\documentclass[12pt,showpacs,preprintnumbers]{revtex4}

\usepackage{graphicx}
\usepackage{dcolumn}
\usepackage{bm}
\usepackage{amssymb}
\usepackage{epsfig}    
\def\beq{\begin{equation}}
\def\eeq{\end{equation}}
\def\ba{\begin{array}}
\def\ea{\end{array}}
\def\bea{\begin{eqnarray}}
\def\eea{\end{eqnarray}}

\def\sqd{\sqrt{2}}

\def\End{\end{document}}

\begin{document}                                                              

\title{New Physics effects on hadronic decay asymmetries of the Top quark.}%

\author{%
{F.~Larios\,$^1$} and
~~{M.A.~P\'erez\,$^2$}  
}
\affiliation{%
\vspace*{2mm} 
$^1$Departamento de F\'{\i}sica Aplicada, 
CINVESTAV-M\'erida, A.P. 73, 97310 M\'erida, Yucat\'an, M\'exico\\
$^2$Departamento de F\'{\i}sica, CINVESTAV, A.P. 14-740, 07000,
M\'exico D.F., M\'exico
}

\begin{abstract}
\hspace*{-0.35cm}

We study some New Physics effects on the hadronic decays
of the Top quark, like $t\to b \bar b c$, and on a
forward-backward-like CP even asymmetry $A_t$ constructed
in such a way that it is zero in the SM. We find that an
anomalous right-handed contribution of the effective $tbW$
vertex may induce an asymmetry $A_t$ of the order of $20 \%$.
A light $W'$ boson with pure right handed couplings $tdW'$
may induce an asymmetry $A_t$ of the same order of magnitude.


\end{abstract}

\maketitle

\setcounter{footnote}{0}
\renewcommand{\thefootnote}{\arabic{footnote}}

\section{Introduction}

The effects of flavor changing neutral currents (FCNC) in processes
involving the top quark have been studied even before the actual
discovery of this quark \cite{rosado}.
While these processes are highly suppressed in the Standard Model (SM),
several FCNC top-quark decays may be enhanced by several orders of
magnitude in scenarios beyond the SM and some of them falling within
the reach of the LHC \cite{roberto}.
FCNC top-quark processes may thus serve as a window to test new
physics effects. Recently, the FCNC contributions to the decay
$t \to b\bar b c$ have been used to identify deviations from the SM
predictions\cite{london}. In particular, it was pointed that four
observables, two CP-even and two CP-odd, that are formulated in
such a way that they are zero in the SM, may produce measurable
values at the LHC and thus may be used to identify new physics
effects.
In this report, we want to focus on a forward-backward-like CP even
asymmetry proposed in Ref.~\cite{london} and compute the predictions
from different New Physics scenarios.  In particular, we are
interested in 1) the contributions from an effective $tbW$ vertex,
and in 2) the contributions induced by the new heavy gauge bosons
introduced to explain the $A_{FB}$ asymmetry measured at the
Tevatron\cite{cao}.  We find that a right-handed ($f_2^R$) $tbW$
coupling may induce a CP-even asymmetry $A_t$ of the order of
$20\%$. On the other hand, a new light $W'$ boson, with mass of
order 180-300 GeV, and with pure right-handed $tdW'$ couplings
may also contribute to $A_t$ with the same order of magnitude.

In any three body decay ($M\to m_1, m_2, m_3$) one can
define three invariant masses $m^2_{ij} =(p_i+p_j)^2$,
that satisfy the constraint
$m^2_{12}+m^2_{23}+m^2_{13}=M^2+m^2_{1}+m^2_{2}+m^2_{3}$.
Only two are independent variables and we can define two
asymmetries depending which $i,j$ pairs we choose.
The asymmetry proposed in Ref.~\cite{london} for the decay
$t\to b\bar b c$ depends on $\rho^2 \equiv (p_b + p_{\bar b})^2$.
However, the SM amplitude squared in the limit $m_b =0$ depends
on a function of $(p_b + p_c)^2$ divided by the $W$-propagator
which depends on $(p_{\bar b} + p_c)^2$.
Therefore, we have chosen $(p_b + p_c)^2$ instead of $\rho^2$
to parametrize the asymmetries of interest.

\section{Asymmetry in the hadronic decay of top}

Let us start by writing the tree level amplitude for the hadronic
decay $t\to b W^+ \to b \bar b c$ of the top quark. For the sake of
generality, we will use the notation $t\to b W^+ \to b \bar d u$
for the initial formula.  Even though our main interest is in
$t\to b W^+ \to b \bar b c$, other decay modes like
$t\to d W^+ \to d \bar b c$,  or $t\to s W^+ \to s\bar s c$
could be addressed in a similar manner.
Besides $u$, $d$ and $s$, we consider $c$ to be massless.  We
allow the $b$ quark mass to be nonzero.  As one would expect,
effects from non-zero $m_b$ are very small and we usually set
$m_b=0$ in our calculations, but we have kept $m_b$ terms in the
initial equations as they can be useful for future studies.
For the propagator of the intermediate $W$ boson we have first
taken the general expression given in Ref.~\cite{atwood} and
then we have simplified it by taking $m_c=0$. The SM amplitude is
\bea
{\cal M} (t\to b u\bar d) &=&
\bar u_b \frac{ig}{\sqd} \gamma_\mu P_L u_t
\bar u_u \frac{ig}{\sqd} V_{ud} \gamma_\nu P_L v_d \;\;
\left( -g^{\mu \nu} + (1+i r_W) \frac{q^\mu q^\nu}{m^2_W}\right) 
G_T^{-1} \, , \label{amplitude} \\
G_T &=& q^2-m^2_W + i q^2 r_W \, ,
\nonumber
\eea
with $r_W \equiv \Gamma_W/m_W$ the width-to-mass ratio of the
$W$ boson.  By momentum conservation $p_t = p_b + p_u + p_d$,
and we define $q\equiv p_u + p_d = p_t - p_b$,
$t\equiv p_b + p_u$ and $\rho = p_b + p_d$.

Let us now consider the particular case $t \to b\bar b c$
with $m_b$ non-zero.
After summing (averaging) over final (initial) spins, and
neglecting terms of order higher than $m_b^2$ the
amplitude squared is given by
\bea
\overline{|{\cal M}_{\rm SM}|^2} (t\to b\bar b c) &=&
\frac{3}{2}  g^4 |V_{cb}|^2 |V_{tb}|^2
M_2(q^2,t^2)|G_T|^{-2}\, ,  \label{smud} \\
M_2(q^2,t^2) &=& t^2 (m_t^2-t^2) + m^2_b \left(
t^2 (2-\frac{m_t^2}{m_W^2}) + m^2_t q^2(m^2_t-q^2)
\frac{1+r^2_W}{4m^4_W} -m^2_t \right) \, .
\nonumber
\eea
Notice that $M^{SM}_2$ depends on $t^2$ in the
numerator and on $q^2$ in the denominator.  One can choose
to apply the constraint $q^2 + t^2 + \rho^2 = m_t^2$ and
change the $t^2$ dependence for a $\rho^2$ dependence as
is done in Ref.~\cite{london}.  As expected, our asymmetry
based on $t^2$ is equivalent to the asymmetry based on $\rho^2$.
As we will see, the definition of the asymmetry involves
finding a particular integration limit that satisfies a cubic
equation. We will find that it is easier to solve the
equation based on $t^2$.

The total width for a three body decay can be written
as\cite{pdg}:
\bea
\Gamma = \frac{1}{(2\pi)^3 32 m_t^3} \int dm_{12}^2
\int dm_{23}^2
\;\;\; \overline{|{\cal M}|^2}.
\label{m12decay}
\eea
Let us now assign $p_1=p_{\bar b}$, $p_2=p_c$ and $p_3=p_b$,
then $m_{12}^2 = q^2$ and $m_{23} = t^2$.  Furthermore, let us
rewrite the integral in terms of dimensionless variables
$x\equiv q^2/m_t^2$ and $y\equiv t^2/m_t^2$
(${\hat m_b}=m_b/m_t$, ${\hat m_W}=m_W/m_t$):
\bea
\Gamma^{SM} (t\to b\bar b c) &=& \frac{m_t}{2^8 \pi^3} \;
\int^{(1-\hat m_b)^2}_{\hat m^2_b} dx \; \int^{y_{max}}_{y_{min}} dy
\;\; \frac{3}{2}  g^4 |V_{cb}|^2 |V_{tb}|^2 \; f^{SM}_2 (x,y)\, .
\label{smwidth}
\eea
Where $f^{SM}_2$ is defined as
\bea
f^{SM}_2 (x,y) &=&
\frac{-y^2 + a y + \frac{1}{4}c_x}{(x-\hat m^2_W)^2 + x^2 r^2_W}
\, ,\;\;\;\;
a\; =\; 1+2 \hat m_b^2 - \frac{\hat m_b^2}{\hat m^2_W}\, ,
\;\;\;\;
c_x \; =\; \hat m_b^2 \left( \frac{1+r^2_W}{m^4_W} -4 \right)\, ,
\nonumber
\eea
and the integration limits are
\bea
y_{max (min)} &=& \hat m_b^2 + \frac{x-\hat m_b^2}{2x}
(1-x-\hat m_b^2) \pm \lambda \frac{x-\hat m_b^2}{2x} \, ,
\nonumber \\
\lambda &=&
\sqrt{1+x^2+\hat m_b^4 - 2x - 2\hat m_b^2 -2x\hat m_b^2}
\, . \nonumber
\eea

Our goal is to define an asymmetry that is zero in the SM
but not necessarily so for other models.  We will split the
integral in $y$ into two equal parts, so that:
\bea
\int_{y_{min}}^{t_x} dy \; (-y^2 + a y + \frac{1}{4}c_x) =
\int_{t_x}^{y_{min}} dy \; (-y^2 + a y + \frac{1}{4}c_x)\, .
\nonumber 
\eea

After solving the integral in the above equation we obtain
a cubic equation for $t_x$
\bea
t^3_x - \frac{3}{2} a t^2_x + 
\frac{3}{4} c_x t_x +\frac{b_x}{4} = 0 \; ,
\nonumber
\eea
where $b_x$ is defined by
\bea
b_x &=& \frac{3}{2} c_x ( y_{max}+y_{min} ) +3a
( y^2_{max}+y^2_{min} ) -2( y^3_{max}+y^3_{min} ) 
\nonumber
\eea

There are three solutions to this equation.  The following one
satisfies that $0\leq t_x \leq 1$ for $0\leq x\leq 1$:
\bea
t_x &=& \frac{1}{2} {\cal R}e \left[a-(1+i\sqrt{3}) z_x^{1/3}
\right] \; , \label{thetx} \\
z_x &=& a^3 -b_x +\frac{3}{2}ac_x+i\sqrt{r_x}
\nonumber \\
r_x &=& (2a^3-b_x)b_x + c_x\left( c_x^2 +3ab_x + 
\frac{3}{4} a^2 c_x \right) \, .\nonumber
\eea
Notice that in the limit $\hat m_b\to 0$, $a\to 1$,
$c_x\to 0$, $y_{min}\to 0$, $y_{max}\to 1-x$,
$b_x \to (1-x)^2 (1+2x)$ and the expression for $z_x$
simplifies greatly: $z_x=1-b_x+i\sqrt{b_x (2-b_x)}$.
It is easy to see that in this limit we end up with a rather
simple analytical expression for $t_x$ in Eq.~(\ref{thetx}).
As mentioned before, choosing the variable $\rho^2$ instead
of $t^2$ leads to a more complicated cubic equation.  Notice
that the authors in Ref.~\cite{london} decided to use a
numerical method to obtain the solution in their study.

The forward-backward-like asymmetry is defined as
\bea
A_t = 
\frac{\int_0^{(1-\hat m_b)^2} dx \int_{t_x}^{y_{max}} dy 
\overline{|{\cal M}|^2} -
\int_0^{(1-\hat m_b)^2} dx \int_{y_{min}}^{t_x} dy 
\overline{|{\cal M}|^2}}
{\int_0^{(1-\hat m_b)^2} dx \int_{t_x}^{y_{max}} dy 
\overline{|{\cal M}|^2} +
\int_0^{(1-\hat m_b)^2} dx \int_{y_{min}}^{t_x} dy 
\overline{|{\cal M}|^2}} \; .
\label{tasymmetry}
\eea
Notice that if NP effects do not change significantly the
value for $BR(t\to b\bar b c)$ in the SM, we can indeed
make an approximation and use the SM value for the
denominator in $A_t$.

The Asymmetry defined in \cite{london} that is based on
$\rho^2$ yields similar results.  For instance, the coupling
constant $X^V_{LR}$ (expected to be of order 1)
associated to a four fermion vector operator yields
$A_\rho = 0.0393 |X^V_{LR}|^2$ (see Eq.~(24) in \cite{london}).
If we consider this coupling and calculate its contribution to
the $A_t$ asymmetry we obtain a value of $0.041 |X^V_{LR}|^2$.

\section{Effective $tbW$ effects on $A_t$}

The first example of NP effects that we want to consider is
based on the effective $tbW$ vertex\cite{chen}:
\bea
\mathcal{L}_{\mathrm tbW} &=&  \frac{g}{\sqrt 2}\, W^-_\mu \, 
\bar b \, \gamma^\mu  \left( f_1^L P_L + f_1^R P_R  \right)\, t
\nonumber \\
&-& \frac{g}{\sqrt 2 M_W} \,
\partial_\nu W^-_\mu \, \bar b \, \sigma^{\mu\nu}
\left( f_2^L P_L + f_2^R P_R \right) \, t \;\;\;+\; h.c.\, ,
\label{tbwvertex} 
\eea
In the SM (tree level) the coupling constants are
$f_1^L = V_{tb} \simeq 1$ and $f_1^R = f_2^R = f_2^L =0$.
The amplitude squared given by this vertex is (see Eq.~\ref{smud})
\bea
M_2 (x,y) &=& -2 {\hat m_b} f_1^L f_1^R x + f_1^L f_2^R
\frac{2}{\hat m_W} x (y-{\hat m_b^2}) - f_1^L f_2^L
\frac{2\hat m_b }{\hat m_W} x(1-y) \nonumber \\
&+& (f_1^R)^2 (x+y-{\hat m_b^2}) (1-x-y) +
(f_2^R)^2 x(y-{\hat m_b^2}) (x+y-{\hat m_b^2})/{\hat m_W^2}
\nonumber \\
&+& (f_2^L)^2 x(1-y) (1-x-y)/{\hat m_W^2}
-f_2^L f_2^R \frac{2\hat m_b^2}{\hat m_W^2} x^2 \; .\nonumber
\eea
The asymmetry in terms of the effective couplings is then
given by
\bea
A_t &=& 0.01 f_1^L f_1^R + 0.04 f_1^L f_2^L + 0.60 f_1^L f_2^R
\nonumber \\
&-& 0.44 (f_1^R)^2 - 1.23 (f_2^L)^2 + 0.78 (f_2^R)^2
+ 0.01 f_2^L f_2^R \, .   \label{atfortbw}
\eea
In principle, we expect the coefficients to assume values
somewhat (maybe much) less than one.  To estimate how large can
$A_t$ become due to effects from the general $tbW$ vertex we
will consider the bounds presented in one recent study based on
$b\to s \gamma$ measurements\cite{misiak}:
$|f^R_1| \leq 2.5 \times 10^{-3}$,
$|f^L_2| \leq 1.3 \times 10^{-3}$ and $|f^R_2| \leq 0.57$.
The potential contributions to the asymmetry are
$A_t \leq 3\times 10^{-5}$, $A_t \leq 5\times 10^{-5}$ and
$A_t \leq 0.34$ respectively.

\subsection{Asymmetry based on $q^2$}

We can define an asymmetry in terms of the $x=q^2/m^2_t$ variable.
Let us simplify formulas by taking $m_b=0$ in this case,
the asymmetry is defined as
\bea
A_q = 
\frac{\int_0^1 dy \int_{q_y}^{1-y}  dx \overline{|{\cal M}|^2}-
\int_0^1 dy \int_0^{q_y} dx \overline{|{\cal M}|^2}}
{\int_0^1 dy \int_{q_y}^{1-y}  dx \overline{|{\cal M}|^2}+
\int_0^1 dy \int_0^{q_y} dx \overline{|{\cal M}|^2}}
\label{qasymmetry}
\eea

The value of $q_y$ is 
\bea
q_y &=& \frac{\hat m_W^2 -\sqrt{(1-y-{\hat m^2_W})^2+r^2_W a^2_y} }
{2-(1-y)(1+r^2_W)/{\hat m^2_W}} \; , \nonumber 
\eea
For some values of $y$, the denominator in $q_y$ approaches zero.
In this region, we can expand $q_y$ in terms of a variable
$\beta \ll 1$:
\bea
\beta &=& \frac{2-(1-y)(1+r^2_W)/{\hat m^2_W}}{(1+r^2_W)^2}
\nonumber \\
q_y &=& \frac{\hat m^2_W}{(1+r^2_W)^2} 
\left( 1-\frac{r^2_W}{2} (\beta + \beta^2 + \cdot \cdot )  \right)
\nonumber 
\eea

The asymmetry in terms of the effective $tbW$ couplings as well
as the four fermion operator of Ref.~\cite{london} is given by
\bea
A_q &=& 0.31 f_1^L f_2^R + 0.01 (f_1^R)^2 + 0.10 (f_2^L)^2
+ 0.13 (f_2^R)^2 + 0.01 |X^V_{LR}|^2    \nonumber
\eea
Comparing with the coefficients in Eq.~(\ref{atfortbw}) we
see that the sensitivity of $A_q$ to NP effects is much lower
than the sensitivity of $A_t$.  Thus, we focus our attention
on the latter.

\section{Asymmetry from models associated with $A_{FB}$}

There are a good number of models proposed in the literature
that attempt to explain the FB asymmetry of $t\bar t$
production at the Tevatron\cite{fbexp}. In many cases the
production process is modified by NP effects that are based
on FCNC couplings\cite{models}.

Let us first refer to three examples presented in Ref.~\cite{cao}.
In particular, we want to consider the cases where there is a
heavy boson involved with a mass of 2 TeV that can contribute
to the $q \bar q \to t \bar t$ process via a t-channel diagram.
One is a neutral vector $Z'$, another is a charged vector $W'$,
and the third one is a neutral $SU(2)$ scalar $S$.  It is possible
that these new bosons contribute to the $t\to b\bar b c$ decay
mode that we are interested.  For instance, the top quark could
make a transition $t\to b {W'}^+ \to b\bar b c$ (i.e. just as
the SM decay but with $W$ being replaced by $W'$).  Unlike the
SM $W$ boson, the $W'$ can couple to both left and right
chiralities and the amplitude squared will have a different form
than Eq.~(\ref{smud}).  Another case is a heavy neutral scalar
$H'$ with strong flavor violating couplings that contributes via
the process $t\to c H' \to c b\bar b$ (or maybe
$t\to c H' \to c b\bar s$).  Similarly the process induced
by the scalar could also stem from a heavy vector boson $Z'$.
For more details we refer to Ref.~\cite{cao}.
Table~\ref{fcboson} shows the contributions to $A_t$ from
each case.

\begin{table}[ht]
\begin{tabular}{|c|c|c|}
\hline model  & couplings & $A_{t}$   
\tabularnewline \hline \hline
$e\; \bar q_d \gamma^{\mu} (f_L P_L + f_R P_R)t {W'}^-_\mu$
& $f_L=2$, $f_R = 20$ & $5\times 10^{-3}$  
\tabularnewline \hline
$y_{ij} \; \bar q_i q_j {H'}$ & $y_{tc}=15$, $y_{bb}=1$ & $0.05$  
\tabularnewline \hline 
$e f_R\; \bar c_L \gamma^{\mu} t_R {Z'}_\mu$
& $f_R = 15$ & $-0.09$  
\tabularnewline \hline 
$\sqd g_s\frac{\kappa_{tcg}}{\Lambda}\;
\bar b_L \sigma^{\mu \nu} T^a t_R G_{\mu \nu}$
& $\frac{\kappa_{tcg}}{\Lambda} \leq \frac{0.06}{\rm TeV}$ & $-0.07$  
\tabularnewline \hline
\end{tabular}
\caption{Asymmetries predicted by FC interactions. The first three
involve new heavy bosons with mass $m_V = 2$ TeV that can give
rise to a $t\bar t$ FB asymmetry $A_{FB} =0.2$\cite{cao}. The
last row is for an effective dimension 5 $tcg$ operator\cite{tcg}
(limits from Tevatron measurements\cite{tcglimits}).
\label{fcboson}}
\end{table}

Another possible FCNC (this one is not used to explain $A_{FB}$)
that we would like to consider involves the gluon field via
a dimension 5 $tcg$ tensor coupling\cite{tcg}.  This coupling
could be generated at loop level and could induce a new single
top production process $gc \to t \to b W^+$ where the top quark
is produced without any additional particles.  Table~\ref{fcboson}
shows the contribution to $A_t$ from this coupling.

It is not the purpose of this work to make a comprehensive
study of NP models and their contribution to the asymmetry $A_t$.
For the results shown so far it has become apparent that there
are several NP scenarios, that are of interest for the Top quark
research program and that could also yield a non-zero $A_t$.
From the results in Table~\ref{fcboson} the heavy $W'$ has
no significant contribution to $A_t$.  However, a light $W'$
with pure right handed FV couplings $tdW'$ \cite{cao2} could
indeed give significant contributions. Table~\ref{wprime} shows
the contributions to $A_t$ for this case.  Notice that the
decay mode is not $t\to b\bar b c$ as before but
$t\to d\bar b c$.
This mode has a negligible width in the SM ($\sim 10^{-8}$ GeV),
so the mere observance of this decay would signal NP effects.
For this decay the denominator in Eq.~(\ref{tasymmetry}) is not
the SM value but the value obtained from the same $W'$ contribution.

\begin{table}[ht]
\begin{tabular}{|c|c|c|c|}
\hline $m_{W'}$(GeV) & $g_R$  & BR($t\to d\bar b c$) & $A_{t}$   
\tabularnewline \hline \hline
$180$ & $1.4$ & $3.7\times 10^{-3}$ & $0.30$  
\tabularnewline \hline 
$200$ & $1.5$ & $2.3\times 10^{-3}$ & $0.11$  
\tabularnewline \hline 
$300$ & $2.0$ & $1.0\times 10^{-3}$ & $-0.13$  
\tabularnewline \hline
\end{tabular}
\caption{Asymmetries predicted by three values of $m_{W'}$
and coupling constant $g_R$ in the case of a light $W'$
\cite{cao2}.  All cases are consistent with $A_{FB} =0.2$.
\label{wprime}}
\end{table}



\noindent
{\bf Acknowledgments}~~~
We thank RedFAE, Conacyt and SNI for support.  We thank
Qing-Hong Cao for providing information on the $tdW'$
coupling.



\end{document}